# Miniaturised control of acidity in multiplexed microreactors

*Divya Balakrishnan [a,b], Wouter Olthuis [b], César Pascual-García [a,*]*

a. Luxembourg Institute of Science and Technology (LIST), 41 Rue du Brill, L-4422 Belvaux, Luxembourg

b. MESA+ Institute, University of Twente, Drienerlolaan 5, 7522 NB Enschede, Netherlands.

**Abstract**

The control of acidity influences the structural assembly of biopolymers that are essential for a wide range of applications. Its miniaturization can increase the speed and the possibilities of combinatorial throughput for their manipulation, similarly to the way that the miniaturization of transistors allows the high throughput of logical operations in microelectronics. Here we present a device containing multiplexed micro-reactors, each one enabling independent electrochemical control of the acidity in ~ 2.5 nL volumes, with a large acidity range in aqueous solutions from pH 3 to 7 and an accuracy of at least 0.4 pH units. The attained pH within each microreactor (with footprints of ~ 0.3 mm$^2$ for each spot) was kept constant for long retention times (~10 minutes) and over repeated cycles >100. The acidity is driven by redox proton exchange reactions, which can be driven at different rates that influence the efficiency of the device in order to achieve more charge exchange (larger acidity range) or better reversibility. By the performance in the acidity control the miniaturisation and the possibility to multiplex paves the way for the control of combinatorial chemistry through pH and acidity controlled reactions.

**Introduction**

Acidity is a parameter that can drive chemical reactions, its control is one of the most important strategies in processes as important as the assembly of DNA[1-4], the control of polymer structures[5,6] or the solid phase synthesis of biopolymers like peptides[7], nucleotides[8] and saccharides[9], which are the basis of combinatorial chemistry. The miniaturized control of these reactions has the potential to provide several advantages: the increase in the kinetics by reducing the distance between the molecules, the possibility to increase the number of simultaneous reactions providing a multiplexed control to achieve high combinatorial throughput with high-density arrays and the reduction of consumption of reagents, minimizing the costs of production. One approach to the miniaturization of chemical reactions controlled by acidity is the classical approach of solid phase synthesis to deliver reagents and immobilize them into a substrate, while applying the same acid treatment into different spots[7,10]. On the other hand, producing the acid locally has the advantage that would allow to introduce reagents in parallel and multiplex the chemical reactions. This strategy decreases greatly the time of high throughput assays. A successful approach to this was driven by Price et al.[11], who used spin casted photo acid generators introduced in a photosensitive resist. With this method they produced arrays with spots of area 50 x 50 $\mu m^2$. A similar method was reported with an array of reactors that used photo acid generators immobilized with other reagents in the reactor volumes for the synthesis of DNA[12]. Here the generated acid was confined in the space of the reactor volume and the photo generating molecules were flushed into the reactors for every chemical reaction. Photoactive molecules were also used to release OH- ions to alter the pH to basic conditions, while controlling enzymatic reactions on DNA[13]. Though photo-activated processes have shown that they can drive reactions in acid and basic conditions in miniaturized size, they require complex equipment and processes.

An alternative approach to generate the acid locally is to use electrochemical redox processes involving proton exchange reactions. Southern and Egeland[14] reported a device consisting of a reaction chamber containing an array of microelectrodes on a glass substrate with an electrolyte containing redox active molecules that could exchange protons with the electrolyte. When a current was applied to the microelectrodes acid was generated in close proximity to them. As the redox reactions can be reused several times, this method has also the advantage that it can be used for several cycles without changing the assembly. However, the fast diffusion of protons limited the range of acidity and the contrast between spots to achieve high reaction yields[15]. To avoid the diffusion of protons, the local control of acidity was improved by implementing a porous substrate on the electrodes and scavenging molecules in the electrolyte[16]. The reactions were confined inside the porous substrate in proximity to the electrodes, and protons would be annihilated outside. While this method improved the local control of acidity to limit the reaction sites into few tenths of micron, the acidity range was still limited. Recently we reported a device to control acidity in nL volumes using a platform that consisted of an electrochemical cell with working, counter and reference electrodes separated by long channels that acted as diffusion barriers[17]. The barriers were designed to avoid the reduction of protons at the counter electrode and to confine the proton concentration in the working electrode cell. Using polymerized 4-Aminothiolphenol (ATP) as redox active molecules functionalized on the electrode surfaces an applied current generated the acid through the oxidation reaction. We reported a binary actuation of the pH in a range from 7 to 1, the stability of the attained pH for longer than 15 minutes and the reversibility of the oxidation and reduction reactions.

Here we present a device able to produce large acidity contrasts in confined microreactors and maintain it for times much longer than what the diffusion of the protons in free space would allow. The new device contains four electrochemical cells with each containing a working, reference, counter electrode cells separated by diffusion barriers. The new reactors use less than 2.5 nL volumes in an area of ~ 0.3 mm². In addition to the binary mode of pH actuation from 7 to 2.6, we also show the quantitative control of pH with an accuracy of ~ 0.4 pH units. Furthermore the multiplexed control of acidity in two of the reactors showing contrast pH changes are presented. The performance of the device in terms of the reversibility of the acidity control and the stability of attained pH are also reported.

**Results and Discussion**

**Design of the chip and description of the setup**

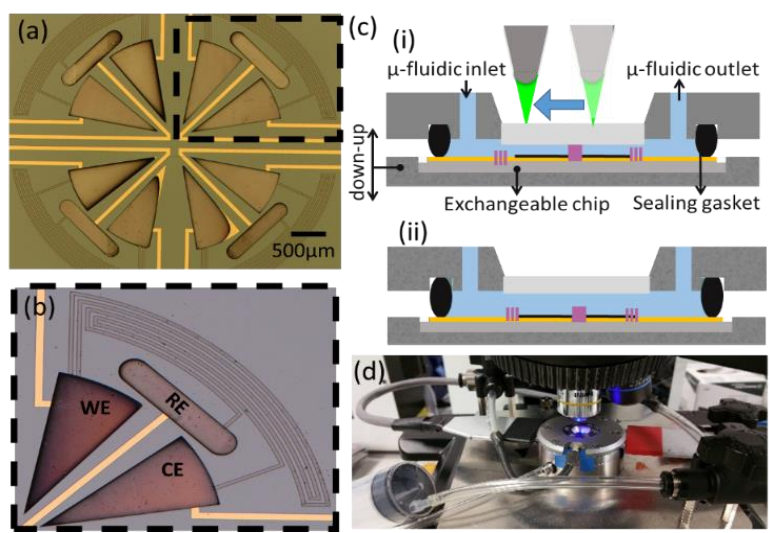

**Figure1:** (a) Microscope image of the multiplexed device chip containing four fully functional electrochemical microreactors. (b) Enlarged view of one the microreactor with WE, RE, CE cells separated by diffusion barriers. (c) Photographic image of the mounted platform. (d) Schematic representation of the microfluidic platform in (i) closed and (ii) open position with movable optical setup on top.

Figure 1 (a) shows the microscope image of the center part of one of the fabricated chips that includes four microreactors. The enlarged view of one of them is also shown in Figure 1 (b) and consists of working, counter and reference electrodes which are confined in each of the different cells. The working principle of each cell is similar to the one in our previous publication[17]. The working electrode cell is separated from the counter and reference electrode ones through two separate long channels that act as diffusion barriers. The electrodes are functionalized with 4 Aminothiolphenol (4ATP). After a polymerization process of these molecules, they dimerize to a redox molecular state[18], which at low applied voltages reversibly exchange protons with the solution thereby changing the pH locally on each of the electrode cell. The long channels were designed to avoid the reduction of protons at the counter electrode. The diffusion of protons along the channel was considered to be one dimensional and the barriers were designed so that the volume of the channel was much smaller than the volume of the electrode cells to avoid the change of acidity in each micro vessel during the diffusion time of protons. The diffusion time of the protons $\tau$ depends on the length of the barrier $L$ and diffusion constant of protons ($D=9\cdot10^{-5} cm^2$/s) and can be calculated as, $\tau=L^2/\pi^2 D$[19]. The channel length of ~ 0.8 cm was designed to provide a diffusion time of 11 minutes for the protons to reach the counter electrode. The width and the height of the barrier were ~25 μm and ~ 7.2 μm respectively. The volume of the electrode cell was ~2.5 nL. Different identical chips were fabricated first with the deposition of electrodes by optical lithography and e-beam evaporation of gold and then the channel barriers for the cells were patterned with optical lithography on a spin coated SU8 epoxy resist. The detailed steps of fabrication were explained in the supporting information figure SI-1. An electrochemical platinisation was performed on the electrode surfaces to increase their surface area. Finally, the electrodes were functionalized with 4ATP to form a self-assembled monolayer of redox active molecules on top of the porous electrodes.

The chip was mounted on a custom-made microfluidic platform schematically represented in Figure 1 (c), the closed position that includes the optical set-up in Figure 1 c (i) and the open position (ii) is shown. The top part of the platform consists of an optical window, microfluidic inlet and outlet and electrical contacts that connect the chip to an external switch box which connects also to a potentiostat. The bottom part also includes a movable piston that allows to move the chip up and down to close and open the cell. The chip was sealed with an O-ring that helps to isolate the liquid from the contact pads during the open position. The platform was mounted under an optical microscope that allowed to monitor the pH using the fluorescence marker carboxy semi-napthorhodafluors (cSNARFs). We used 1 M concentration of Potassium chloride solution with 0.5 µM cSNARF as electrolyte. For pH control measurements, the microscope was moved to the respective electrode cells of the different microreactors to follow the changes in the spectra of the fluorescence marker that arises due to the redox active reactions at the electrode surface. Figure 1 (d) shows the image of the mounted platform. More details of the experimental setup were described in our previous works with chips where the reactors were not miniaturized[17, 20]. When the chip was closed by the glass window from the top, the epoxy resist ensures the isolation of the electrode cells and channels. On the current chip designed we used a circular arrangement of electrodes and channels at the center so that they occupy most of the flat available surface of the spin coated resist. Additionally the concentric arrangement of channels could relax the tensions producing less cracks on the processed samples.

**Control of acidity in one microreactor.**

The acidity in the cells was controlled by the reversible redox reactions of the 4ATP that exchange protons with the electrolyte. The redox active states of 4ATP were created by polymerizing electrochemically the self-assembled monolayer[21] with the chip in open position to reduce the external resistivity existing when the electrodes are connected only through the diffusion barriers (closed cell configuration). Following the electropolymerization, the platform was flushed with fresh electrolyte and placed in closed position to constrain the volume for pH control measurements. Each cell on the chip was connected to a potentiostat channel with three-electrode configuration. Cyclic voltammetry (CV) was used to drive the redox reactions on the polymerized 4ATP layer. The microscope was focused on the WE to monitor the acidity changes through the fluorescence spectrum during CV measurements. The pH inside the electrode cell was monitored by the SNARF signal which has two peaks, the one at 590 nm is dominant in acidic pH and the other at 655 nm is dominant in basic pH. The relative intensity of the two peaks can be used to calibrate the pH. Due to the miniaturized volume and low concentration of SNARF we used an excitation wavelength at 565 close to the absorbance of SNARF molecule and to avoid the stray light we used a filter around 600 nm.

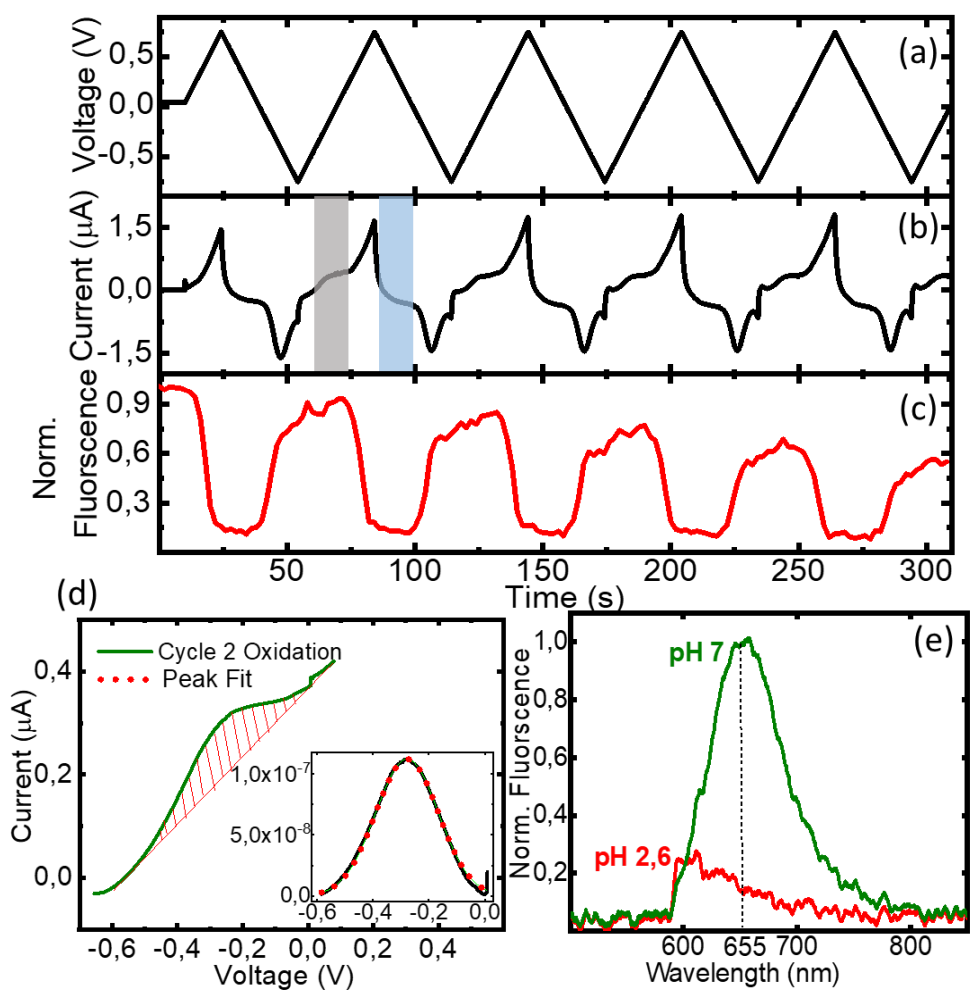

**Figure 2:** The graph shows (a) the voltage, (b) the current and (c) the normalized intensity of the fluorescence marker plotted against time. (d) Current is plotted against voltage for the cycle 2 (data corresponds to the shaded region in (b)). The inset shows the Gaussian fitting of the peak in red dotted plot and the subtracted data of current vs voltage is shown in black. (d) SNARF fluorescence spectrum showing the maximum and minimum intensity peaks corresponding to the pH value 7 and 2,6 respectively.

Then we followed the pH variations with the intensity of the peak of SNARF at 655 nm. The correspondence between the 655 nm fluorescence peak intensity normalized at the initial pH 7.4 and the real pH is shown in the calibration plot SI figure 4. To follow the fluorescence each spectrum was integrated for 1 second during the CV.

Figure 2 (a), (b) and (c) show the bias voltage, the corresponding current and the 655nm peak fluorescence normalized to the peak intensity at pH 7.4, respectively, plotted against time during 5 cycles. The voltage range used was -0.75 V to +0.75 V at a rate of 50 mV/s. The positive voltage sweep from 0 V to 0.5 V gives rise to an oxidation peak like the one shown in figure 2 (d), which corresponds to the grey shadowed region shown in the second cycle of figure 2 (b). At the end of the oxidation peak a decrease in the fluorescence intensity is observed in figure 2 (c), which shows the decrease of the pH in the cell. The negative voltage sweep from 0 to -0.5 V produces a reduction peak shown in the region shadowed in blue in figure 2 (b). At the end of this peak, an increase in the fluorescence intensity was observed in (c) that shows that the neutral pH was recovered. This cycle repeated successively showing reversible changes on the fluorescence intensity driven by the reversible redox states of the 4ATP molecule that releases protons during oxidation reaction and recovers them from the electrolyte during reduction reaction thereby controlling the acidity in the working electrode cell. In figure 2 (b), we also observed a small peak before the main oxidation, which was detected in the successive cycles. The origin of this different oxidation could be attributed to the underlying Au in the electrode due to incomplete platinisation in the electrodes which could produce an oxidation of the 4ATP with slightly different potentials and that we observed only in some samples where the electrodes were not completely platinized. The pH attained in the working electrode cell could be followed until the detection limit of the SNARF maker at pH 5 (see calibration plot shown in figure SI-2). Below this limit the 655 nm peak completely quenches and no more changes in the intensity of the 655 nm peak with respect to the pH was observed. In our CV experiment we started with the electrolyte that was prepared with KCl at pH 7. The minimum pH beyond the range of the SNARF marker (figure 2 (e)) could be estimated calculating the total charge exchanged by the electrons in the cell ($Q$) and attributing that charge exchange to redox reactions producing only protons. As the pH change is more than two units, as monitored by the SNARF, the final concentration exceeds largely the initial one at neutral pH, and as the measuring time is much smaller than the diffusion time of protons through the diffusion barriers The pH could be calculated converting $Q$ into number of protons and retrieving the concentration using the volume in the cell:

$$pH = -log\left[\frac{Q}{F.V_{cell}}\right] \quad \text{Eq. (1)}$$

where $F$ is the Faraday constant and $V_{cell}$ is the volume of the working electrode cell (~ 2.5 nL). The charge $Q$ was obtained from the integrated area under the oxidation peak and the scan rate. In figure 2 (d) the oxidation peak for cycle 2 indicating the area under the peak (shaded region in figure 2 (b)) is shown along with the inset presenting the Gaussian fitting and the results from the fitting are tabulated in the Table 1. The minimum pH attained in the cell resulting from the equation (1) was calculated as 2.6.

*Table 1 Gaussian fitting and the results from the fitting are tabulated in the table T1*

| Peak type | Peak area (C) | Max height (A) | FWHM (V) |
|---|---|---|---|
| Gaussian | 6.32E-7 | 1.12E-7 | 0.27 |

**Quantitative control and Retention time of the acidity**

An accurate control of the acidity in the cell would allow the possibility to orthogonally control chemical reactions specific to different pH values. In our device we tested quantitative control of the acidity by using pulse voltammetry as shown in Figure 3 (a) and (b). Different voltages with pulses of 20 s were applied while monitoring the pH through the fluorescence intensity Figure 3 (a). After the first pulse of 0.8 V was applied we decreased the amplitude in steps of 0.05 V until reaching 0.55 V. Figure 3 (b) shows the normalized fluorescence intensity with respect to the different voltage pulses plotted against the time. After the first voltage pulse at 0.8 V the fluorescence intensity decreased to 0.1 while at 0 V the initial fluorescence intensity was recovered. A similar response was also observed during the second voltage pulse at 0.75 V, probably because the pH marker was out of range. For the next voltage pulse at 0.7 V the fluorescence intensity was slightly higher than that of the previous pulse showing a control of the pH. At 0.65 V the intensity was higher than the previous pulse at 0.7 V and this process was repeated for the next consecutive pulses too. The corresponding minimum pH values were denoted in figure 3 (b), showing that in addition to the binary mode of pH actuation (7 to 2.6), we were also able to demonstrate quantitative control over the acidity concentration with ΔpH of ~0.4 units.

We tested the stability in terms of retention time of the attained pH in the electrode cell by applying a voltage pulse from -0.15 V to 1.2 V for 3 cycles to observe a reversible pH actuation and then we opened the circuit at 1.2 V. The change in the fluorescence intensity was monitored for 10 minutes, which is similar to the calculated diffusion time of the protons through the barrier (11 minutes). Figure 3 (a) shows the applied voltage pulse and (b) the fluorescence spectra plotted against time. In the third voltage pulse at 1.2 V the fluorescence intensity was completely quenched. Until 300 seconds (5 minutes) we observed no noticeable change in the fluorescence intensity. After five minutes a slight increase in the intensity was observed. At 600 seconds (10 minutes) the pH maintained in the cell was < 5 as it was obtained from the calibration plot. These measurements show that the design of the device along with the model of diffusion barrier is compatible with miniaturization and the ability to sustain the acidic conditions for 5 minutes already opens applications to the occurrence of different chemical processes.

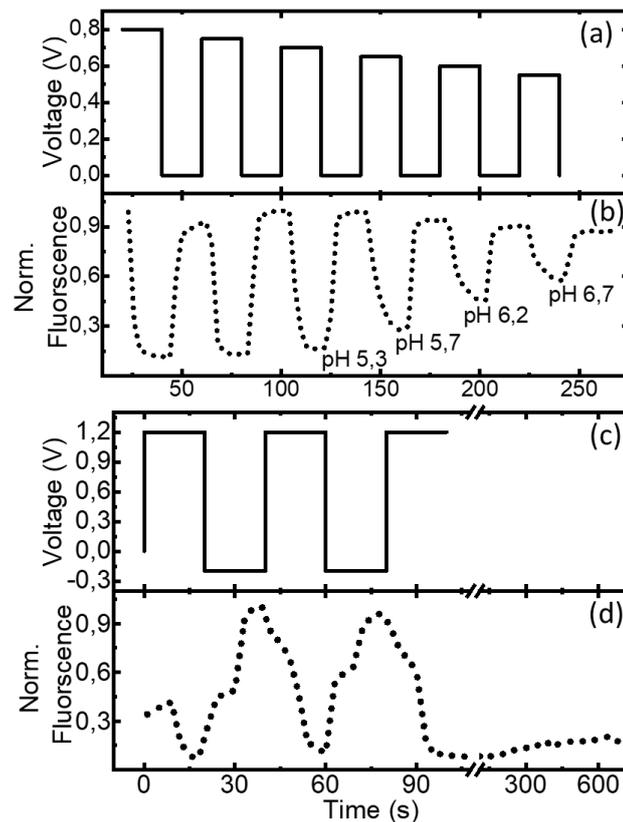

**Figure 3**: The graph shows (a) the voltage pulse and (b) the corresponding fluorescence spectra with different pH values plotted against time. (c) The graph shows the voltage pulse and the corresponding fluorescence spectra plotted against time.

**Reversibility tested over 100 cycles**

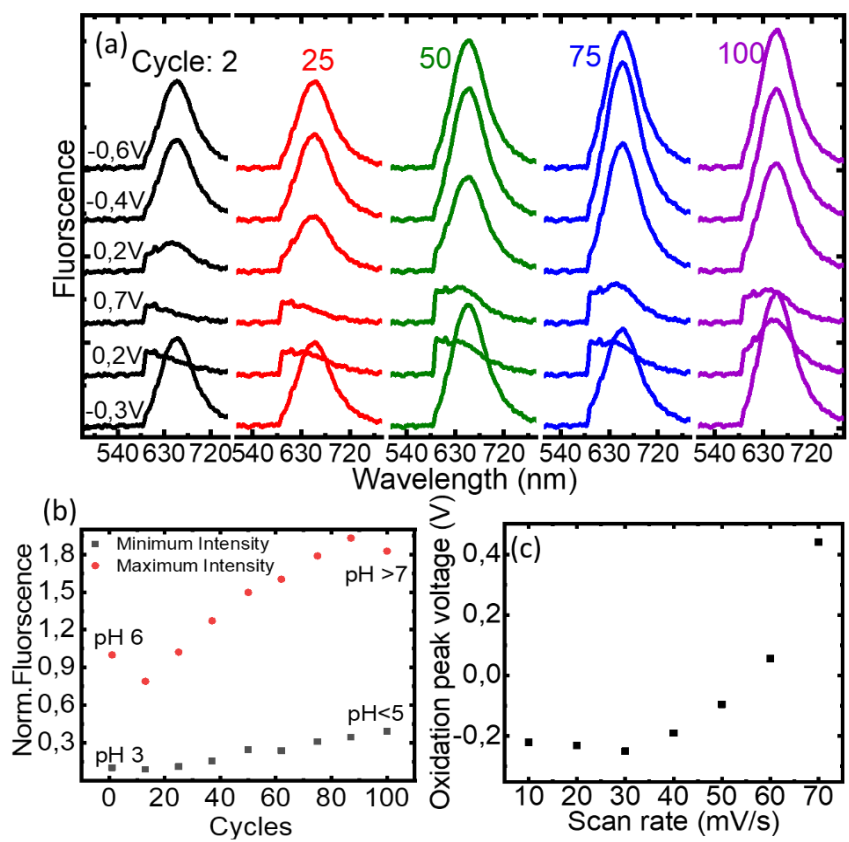

**Figure 4** (a) SNARF fluorescence spectra plotted for the CV cycles 2, 25, 50, 75, 100 in the voltage range from -0,75 V to +0,75V with an offset in Y axis. (b) shows the plot of the CV cycles against the normalized fluorescence intensity of the snarf spectra indicating the maximum and minimum intensity. (c) The graph shows the scan rate plotted against the voltage at the oxidation peak.

We studied the reversibility of acid control monitoring the CV during 100 cycles with the fluorescence marker. The voltage was swept from -0.6 V to 0.7 V at a scan rate of 100 mV/s and the fluorescence spectra were recorded with integration times of 1 second each. Figure 4 (a) shows the spectra corresponding to cycles 2 (black), 25 (red), 50 (green), 75 (blue) and 100 (purple) at different voltages plotted with a vertical offset as indicated in the graph. All the spectra reflect the behavior of pH already explained depending on the oxidation/reduction of the 4ATP. Initially, in cycle 2 (in black) during the voltage sweep at -0.6 V (top spectrum) the fluorescence peak at 655 nm corresponds to the neutral pH in the electrode cell. Towards positive voltages around 0.7 V (fourth spectrum) the 655 nm peak was quenched due to the increase of acidity made by the 4ATP oxidation and at -0.3 V (bottom spectrum) the increase in peak intensity witnesses the recovering of more basic conditions. A similar behaviour in each of the cycles 2, 25, 50, 75, 100 was observed where the peak at 655 nm appeared and quenched indicating neutral and acidic pH conditions in the cell respectively. Figure 4 (b) shows the cycle dependence of the maximum and minimum peak intensities of the 655 nm plotted in red and black, respectively. After 25 cycles (40 minutes) we observed an increase of both intensities, which reflects a rise in the overall acidity of the system. After flushing the cell with fresh electrolyte, we were able to recover the same intensities repeating the experiment with the same chip (data not shown). This effect of progressive increase of basic conditions obtained with the closed cell could be attributed to the leakages of the protons, and/or to defects in the 4ATP functionalization at the electrode surface which would results in holes in the electrode interface where direct exchange of electrons with protons to reduce them into Hydrogen gas ($H_2$) could occur during the negative voltage bias. Both effects can decrease the overall proton concentration in the cells resulting in an increase in the basicity in the microreactor. Nevertheless, the minimum pH attained in the cell after 100 cycles was still under 5 from our calibration.

The kinetics of the redox reactions was also monitored with CV cycles at different scan rates from 10 mV/s to 70 mV/s. We observed a shift in oxidation peak potential. In Figure 4 (c) the scan rate is plotted against the voltage of the oxidation peak. The oxidation peak voltage was around -0.2V for upto 40mV/s. Then with increasing scan rates, the oxidation peak voltage shifted to higher voltages, which may be because those rates did not allow enough time for the redox reaction to occur. This indicates that an increase acidity control with lower bias could be used until 30 mV/s, although that would increase largely the control of multiple cycles, and the reversibility shown in figure 4 (a) provides already a pH contrast of at least two units for more than 100 cycles.

**Multiplexed control of acidity in two cells with different phase**

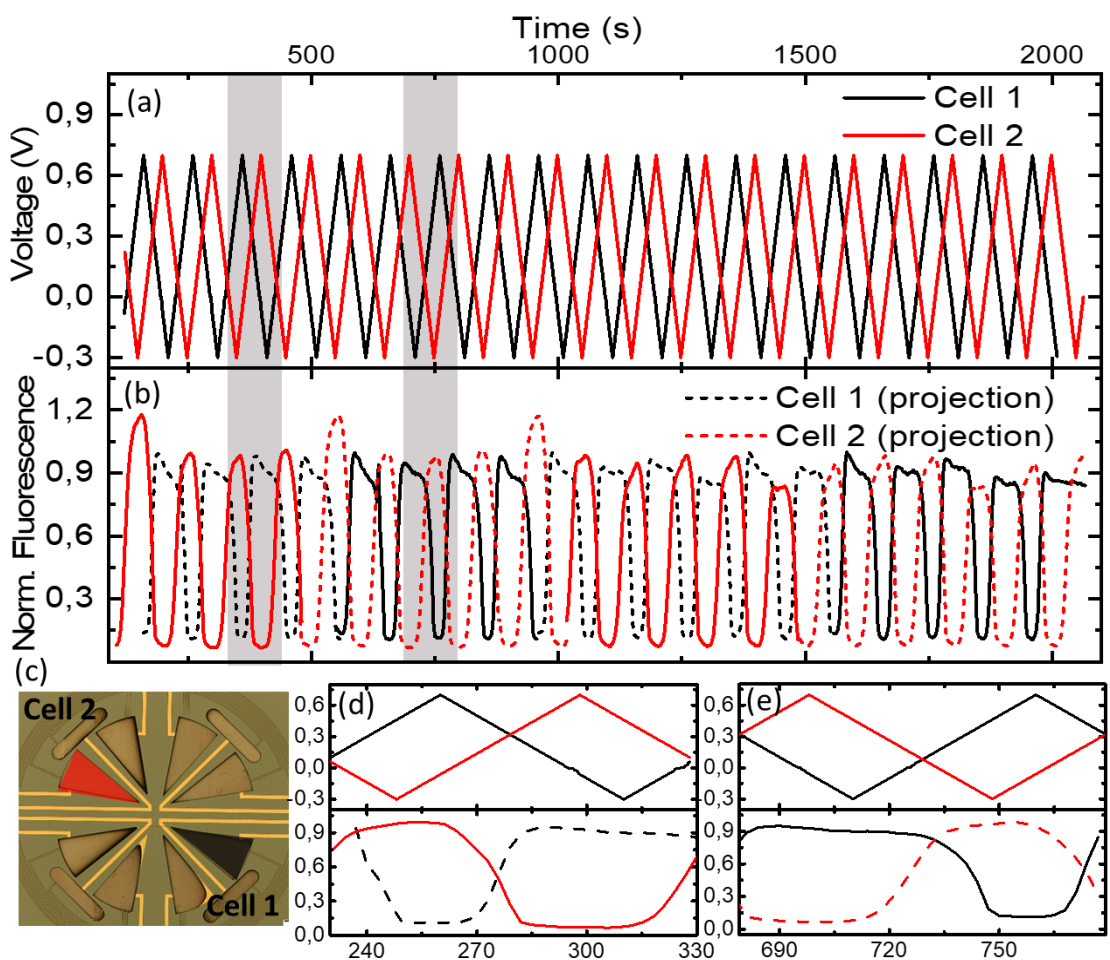

**Figure 5**- shows multiplexed control of acidity in cell 1 and cell 2. (a) The voltage applied in cell 1 and cell 2 is shown in black and red respectively, and (b) the corresponding fluorescence spectra is plotted against time. The dotted plots of cell 1 and cell 2 represent the projection of the spectra (c) Microscope image of the chip highlighting the WE of cell 1 and cell 2. (d) and (e) shows the enlarged view of the dotted area in (a, b).

The chip with multiple cells was designed to allow simultaneous control of acidity in different cells. To control this feature two channels of the potentiostat were connected and we adjusted the CV cycles to apply the voltage out of phase introducing delay time at the start of the CV of channel 2. We used a voltage range of -0.3 V to 0.7 V at a scan rate of 20 mV/s in both channels. To measure the fluorescence signal in each cell using our single channel spectrometer we alternated the position of the cells two times in each cell every 5 cycles to complete twenty cycles. Figure 5 shows (a) the applied voltage and (b) the corresponding fluorescence intensity plotted against time in cells 1 and 2 in black and red, respectively. The WE cells 1 and 2 are shadowed in black and red respectively in Figure 5 (c). To visualize the behaviour of acidity, the data from the previous 5 cycles of the two cells was projected in dotted lines for the cycles where the microscope objective was on the other cell, since simultaneous measurement of fluorescence intensity at the two cells was not possible. Both cells 1 and 2 (plotted in black and red, respectively) show the behaviour already described with oxidation and reduction at ~0.7 V and 0.1 V respectively, and the fluorescence following the acidic behaviour of the redox reactions. The fluorescence is shown to be out of phase on both cells, following the respective bias actuation. It is also remarkable that the projected fluorescence plotted in dotted lines for both cells shows very good continuity with the real fluorescence, which indicates that during the time the fluorescence is not monitored in each cell, the acidity follows the expected behaviour. The enlarged views of the CV cycles in the shaded areas in figures 2 a and b are shown in (d) and (e), respectively. The changes in fluorescence intensity with respect to the applied voltage bias could be seen along with the contrast fluorescence changes in the other cell which is denoted as a projection. For 20 cycles alternating pH conditions acidic – neutral was demonstrated in cell 1 and cell 2 maintaining the contrast conditions. No potential leakage was observed. As conclusion we drove the multiplexed control of the acidity in these two miniaturized cells in an independent way.

**Conclusions and Outlook**

We fabricated a miniaturized device that holds four microreactors to control the acidity in miniaturized cells. The control of pH was demonstrated using the electrochemical reactions of redox active molecules functionalized on the surface of the electrodes. The microreactor was designed to confine the protons in electrode cells while the reduction during negative biases was avoided using diffusion barriers separating working and counter electrodes. A fluorescence marker in the electrolyte was used to track the pH. We were able to control the acidity in the individual microreactor from pH 7 to complete quenching of the marker at 5. The minimum pH attained in the cell was 2.6 as was calculated from the charge transfer of the polymerized 4ATP molecules. We demonstrated that regardless the miniaturization, the pH was controlled similarly to our previous article, which demonstrates that the acidity control is independent of the footprint of the microreactors. Further miniaturization of the device would be possible thanks to our design. We showed a quantitative control of pH with an accuracy 0.4 pH units varying the amplitude of the applied potential pulses. The microreactor retains a stable pH in the electrode cell for 10 minutes by the confinement of protons, thanks to the tailored diffusion barriers. This time control would allow to drive chemical reactions such as synthesis of bio-polymers, that we expect to occur faster than in macroscopic vessels because the diffusion of reagents within this cell is limited with the miniaturized volumes. The reversibility of the acidity control was tested over 100 cycles. The minimum pH corresponding to the $100^{th}$ cycle was still under <5. This shows that the system has good control over the reversibility providing a dynamic acidity control and well suited for combinatorial chemistry applications that allows the assembly of molecules on the same spot. For example one of the current challenges of combinatorial chemistry is the peptide synthesis. Current methods use 50 % concentration of triflouroacetic acid to remove the protecting groups in macroscale reactors. The pH corresponding to this concentration is ~ 5.9 (more information is provided in the supporting information). As the generation of acid through the redox reactions of 4Aminothiolphenol occur also in organic solvents (data not shown), our device has the potential to implement electrochemically controlled synthesis. The multiplexed control of acidity was demonstrated in two cells on the same chip using simultaneous CV measurement that are out of phase to maintain different pHs without any crosstalk. This shows that our device could be used for combinatorial chemistry applications where the regulation of the large acidity range would enhance the throughput and the yield of chemical reactions.

In summary, we have demonstrated to the best to our knowledge, the largest control of the acidity in terms of acidity range and retention time, driven by electrochemical means in cells miniaturized in the range of hundred microns. We also showed that the pH could be independently activated in different cells. We believe that the key features of our device large pH range, quantitative control, reversibility and multiplexing could be used to control chemical reactions with increased throughput assuring stability of acid contrast between nearby spots.

**Experimental Section**

**Fabrication of the chip**

The chip was fabricated using optical lithography. A maskless aligner – MLA 150 from Heidelberg Instruments was used to pattern the designs. The series of experimental steps involved in the fabrication process are detailed in the supporting information. The substrate was spincoated with optical resist and design of electrodes with contact pads are exposed and developed. Later 5 nm of Titanium and 50 nm of Gold was evaporated using a e-beam evaporator. The chip was placed in acetone for lift-off. Then the chip was spin coated with SU8 3010 epoxy resist to make the second layer. The design of the diffusion barriers and the electrodes are exposed on the chip using the maskless aligner. The chip was then developed and hard baked for 10 minutes. The electrodes are electrochemically platinized [17]. The chip was later cleaned in a UV ozone cleaner for 30 minutes and functionalized with 0.5 mM concentration of 4 Aminothiolphenol in absolute ethanol for 24 hours.

## Electropolymerization and acidity control on the platform

The cell was placed in open position and through the switch box the electrodes are connected to the potentiostat (Pstat 1) by three electrode configuration. A CV program was applied in the range of -0.25 V to + 0.7 V at a scan rate of 50 mV/s for 3 cycles.

Multiplexing experiments were performed in cell closed position with the first potentiostat (Pstat 1) connected to cell 1 of the chip and the second potentiostat (Pstat 2) connected to cell 2 of the chip through the switch box. A CV program was applied between -0.3 V to -0.65 V at a scan rate of 20 mV/s for 20 cycles. The microscope was objective was focused at cell 2 working electrode for the cycles 1 to 5 and 11 to 15. The objective was focused at cell 1 working electrode during the cycles 6 to 10 and 16 to 20.

## Chemicals and Instrumentation

4 Aminothiolphenol, Potassium chloride and absolute ethanol were purchased from Sigma Aldrich. Carboxy semi-napthorhodafluors was purchased from Molecular Probes Inc. For electrolyte preparation and cleaning purposes Millipore filtered water was used. For the electropolymerization and pH control experiments Solatron Modulab XM Pstat 1mS/s potentiostat (Pstat 1) was used. For multiplexing experiments, Gamry Potentiostat REF600-05104 (Pstat 2) was used.

## Conflict of interest

The authors declare no conflict of interest.

## Acknowledgements

This research work was financed by the FNR ATTRACT project 5718158 NANOpH.

## Author contributions

The manuscript was written through contributions of all authors. All authors have given approval to the final version of the manuscript.

# References

1. Xiaolian, G, Erdogan, G, Xiaochuan, Z. In situ synthesis of oligonucleotide microarrays, Biopolymers, 2004, doi: 10.1002/bip.20005

2. Liu, D. S, Balasubramanian, S. A proton fuelled DNA nanomachine, Angew. Chem., Int. Ed. 2003, 42, 5734-5736.

3. Liedl, T., Simmel, F. C. Switching the conformation of a DNA molecule with a chemical oscillator, Nano Lett. 2005, 5, 1894-1898.

4. Liu, D. S., Bruckbauer, A., Abell, C., Balasubramanian, S., Kang, D. J., Klenerman, D., Zhou, D. A Reversible pH-Driven DNA Nanoswitch Array, J. Am. Chem. Soc. 2006, 128, 2067–2071.

5. Shu, C. F.; Wrighton, M. S. J. Phys. Chem. 1988, 92, 5221–5229.

6. Niwa, M.; Mori, T.; Higashi, N. Macromolecules, 1995, 28, 7770-7774.

7. Merrifield R. B. Solid Phase Peptide Synthesis. I. The Synthesis of a Tetrapeptide. J. Am. Chem. Soc. 1962, 85 (14), 2149-2154.

8. Amodio A, Adedeji AF, Castronovo M, Franco E, Ricci F. pH-Controlled Assembly of DNA Tiles. J Am Chem Soc. 2016, 138(39):12735–12738. doi:10.1021/jacs.6b07676.

9. Sears, P., Wong, C.,Toward Automated Synthesis of Oligosaccharides and Glycoproteins, Science, 2001, 2344-2350.

10. Frank R., The SPOT-synthesis technique Synthetic peptide arrays on membrane supports--principles and applications. J Immunol Methods, 2002, 20021: 13-26.

11. Price, J. V., On silico peptide microarrays for high-resolution mapping of antibody epitopes and diverse protein-protein interactions. Nature Med. 2012, 18, 1434-1440.

12. Zhou, X et al., Microfluidic PicoArray synthesis of oligodeoxynucleotides and simultaneous assembling of multiple DNA sequences." Nucleic acids research vol. 32, 2004, 18 5409-17, doi:10.1093/nar/gkh879

13. Elbaz, J, Wang. F, & Remacle. F. Willner, I., pH-Programmable DNA Logic Arrays Powered by Modular DNAzyme Libraries. Nano letters. 2012, 12. 10.1021/nl300051g.

14. Egeland, R. D, Southern, E. M. Electrochemically directed synthesis of oligonucleotides for DNA microarray fabrication. Nucleic Acids Res. 2005, 33, 125.

15. Karl, M., Andy, M., Michael, S., Kilian, D. The Removal of the t-BOC Group by Electrochemically Generated Acid and Use of an Addressable Electrode Array for Peptide Synthesis. J Comb Chem. 2005, 7(5), 637-640.

16. Maurer, K. Electrochemically Generated Acid and Its Containment to 100 Micron Reaction Areas for the Production of DNA Microarrays. PLOS one, e34, 2006, 3903-3908.

17. Balakrishnan, D., Lamblin, G., Thomann, J.S., van den Berg, A., Olthuis, W., Pascual Garcia, C. Electrochemical control of pH in nano-litre volumes. Nano Lett. 2018, DOI: 10.1021/acs.nanolett.7b05054.

18. Hayes, W. A. & Shannon, C. Electrochemistry of surface confined mixed monolayers of 4-Aminothiolphenol and thiolphenol on Au. Langmuir 12, 3688–3694 (1996).

19. Clément, N.; Nishiguchi, K.; Dufreche, J. F.; Guerin, D.; Fujiwara, A.; Vuillaume, D. Nano Lett. 2013, 13, 3903−3908.

20. Balakrishnan, D., Gerard, M., Girod, S., Frari, D.D., Olthuis, W., Pascual Garcia, C. Redox active polymer a a pH actuator on a Re-sealable microfluidic platform. J.Materials Sci.Eng. 2018, DOI: 10.4172/2169-0022.1000456.

21. Balakrishnan, D., Lamblin, G., Thomann, J. S., Guillot, J., Duday, D.; van den Berg, A., Olthuis, W., Pascual-Garcia, C., Influence of polymerisation on the reversibility of low-energy proton exchange reactions by Para-Aminothiolphenol, Sci. Rep., 2017, 7, 15401.